\documentclass[sigconf]{acmart}

\AtBeginDocument{%
  \providecommand\BibTeX{{%
    \normalfont B\kern-0.5em{\scshape i\kern-0.25em b}\kern-0.8em\TeX}}}

\copyrightyear{2021} 
\acmYear{2021} 
\setcopyright{rightsretained} 
\acmConference[EAAMO '21]{Equity and Access in Algorithms, Mechanisms, and Optimization}{October 5--9, 2021}{--, NY, USA}
\acmBooktitle{Equity and Access in Algorithms, Mechanisms, and Optimization (EAAMO '21), October 5--9, 2021, --, NY, USA}
\acmDOI{10.1145/3465416.3483304}
\acmISBN{978-1-4503-8553-4/21/10}

\usepackage{subfig}

\usepackage{enumitem}

\newcommand{\al}{\alpha}
\newcommand{\E}{\mathbb{E}}
\newcommand{\I}{\mathbb{I}}

\newcommand{\cF}{\overline{F}}
\newcommand{\bx}{\textbf{x}}

\DeclareMathOperator*{\argmax}{arg\,max}

\begin{document}

\title{Project 412Connect: Bridging Students and Communities}

\author{Alex DiChristofano}
\authornote{Order is alphabetical.}
\email{a.dichristofano@wustl.edu}
\orcid{0000-0001-5458-6105}
\affiliation{%
  \institution{Division of Computational \& Data Sciences, Washington University in St. Louis}
  \city{St. Louis}
  \state{Missouri}
  \country{USA}
  \postcode{63130}
}

\author{Michael L. Hamilton}
\authornotemark[1]
\email{mhamilton@katz.pitt.edu}
\orcid{0000-0002-4671-395X}
\affiliation{%
  \institution{Katz Graduate School of Business, University of Pittsburgh}
  \city{Pittsburgh}
  \state{Pennsylvania}
  \country{USA}
  \postcode{15260}
}

\author{Sera Linardi}
\authornotemark[1]
\email{linardi@pitt.edu}
\orcid{0000-0002-8085-1555}
\affiliation{%
  \institution{Graduate School of Public and International Affairs (GSPIA), University of Pittsburgh}
  \city{Pittsburgh}
  \state{Pennsylvania}
  \country{USA}
  \postcode{15260}
}

\author{Mara F. McCloud}
\authornotemark[1]
\email{mfm53@pitt.edu}
\orcid{0000-0002-9229-9737}
\affiliation{%
  \institution{Graduate School of Public and International Affairs (GSPIA), University of Pittsburgh}
  \city{Pittsburgh}
  \state{Pennsylvania}
  \country{USA}
  \postcode{15260}
}

\renewcommand{\shortauthors}{DiChristofano, Hamilton, Linardi, and McCloud}

\begin{abstract}
In this work, we describe some of the challenges Black-owned businesses face in the United States and specifically in the city of Pittsburgh. Taking into account local dynamics and the communicated desires of Black-owned businesses in the Pittsburgh region, we determine that university students represent an under-utilized market for these businesses. We investigate the root causes for this inefficiency and design and implement a platform, 412Connect (\url{https://www.412connect.org/}), to increase online support for Pittsburgh Black-owned businesses from students in the Pittsburgh university community. 

The site operates by coordinating interactions between student users and participating businesses via targeted recommendations. We describe the project from its conception, paying special attention to our motivation and design choices. These choices are aided by two simple models for badge design and recommendation systems that may be of theoretical interest. Along the way, we highlight challenges and lessons from coordinating a grassroots volunteer project working in conjunction with community partners and the opportunities and pitfalls of engaged scholarship. 
\end{abstract}



\maketitle

\section{Introduction}\label{sec:intro}

Small businesses are engines of economic mobility. Self-employment has been shown to lead to wealth accumulation four times that of those who are not self-employed \citep{Misera2020}. Despite this, Black-owned businesses face a number of systemic challenges that can inhibit their growth and success. These challenges include lack of access to capital, mentorship networks, and business opportunities, which can limit owners' ability to weather market disruptions \citep{House2018}. Further, historically Black-owned businesses are often geographically segregated from areas of high economic activity, and target markets have insufficient information about them \citep{reardon2008geographic,rutan2018lingering}. These structural inequalities have been further exacerbated by the COVID-19 pandemic, leading to disproportionate economic slowdowns and closures \citep{Mills2020}.

\begin{figure}[ht]
\centering
\includegraphics[width=.95\linewidth]{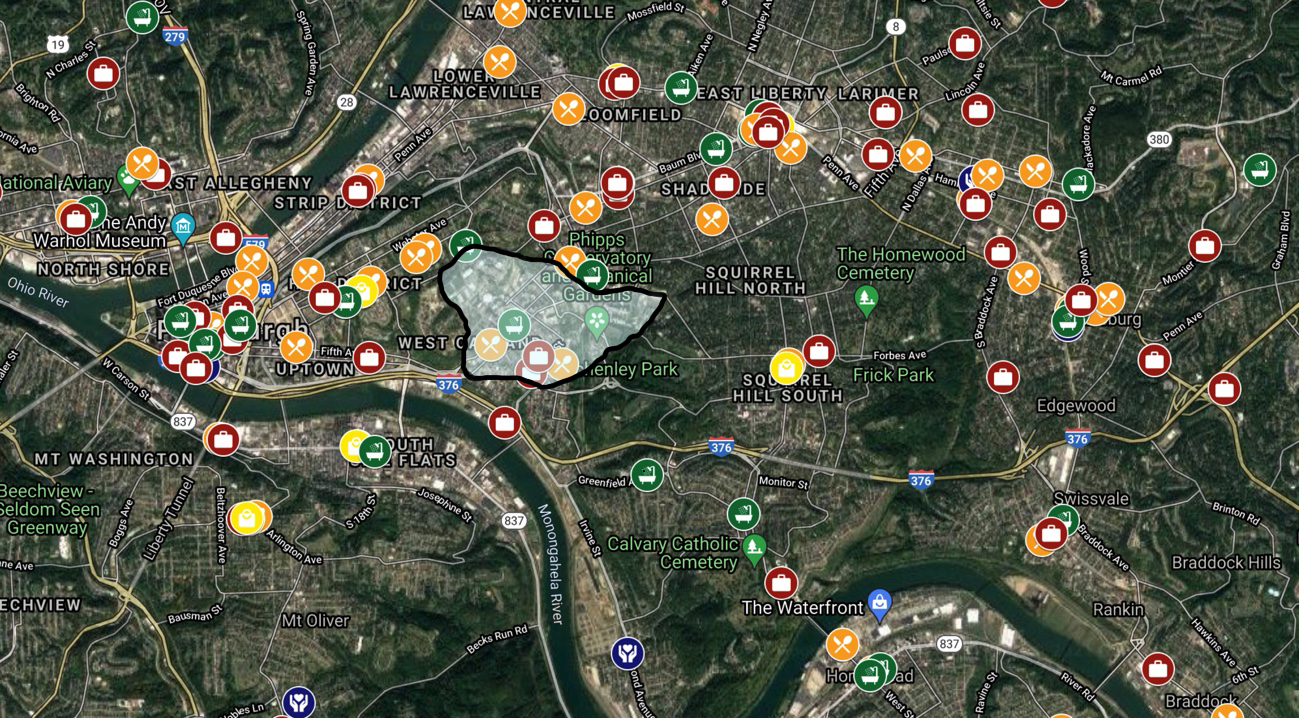}
\caption{Black Owned Businesses in Pittsburgh. {\rm Depicted is a satellite map of the city of Pittsburgh. The icons on the map represent Black-owned businesses in service (red), food (orange), shopping (yellow), and beauty (green) sectors. The shaded section in the middle of the map encompasses the greater university area inside the neighborhood of Oakland, where the majority of students at the University of Pittsburgh reside. This map was created in collaboration with our community partners using Google My Maps. Map Data: \textcopyright 2021, Google. Imagery: \textcopyright 2021, Landsat/Copernicus, Maxar Technologies, Sanborn, U.S. Geological Survey, USDA Farm Service Agency.}}
\label{fig:Black-owned business_map}
\Description[The lack of Black-owned businesses in Oakland.]{Clusters of Black-owned businesses exist throughout Pittsburgh, but there are comparatively few in the university community of Oakland.}
\end{figure}

The challenges Black-owned businesses face nationally also play out locally in the city of Pittsburgh, the community of focus in this work. According to the 2019 American Community Survey (ACS), while the median household income in Pittsburgh was \$62,638 and the median household income of households with white householders was \$65,969, the median household income of households with Black householders was \$35,974 \citep{ACS2019, ACS2019White, ACS2019Black}. Further, the neighborhoods of Pittsburgh are historically segregated by race \citep{reardon2008geographic}, with damaging effects of redlining persisting in Pittsburgh neighborhoods and fostering poverty and vacancy among minority groups in redlined areas \citep{rutan2018lingering}. This is reinforced by a local community culture of neighborhood-based mindsets in which residents conduct most of their activities within their neighborhoods, a disparity especially evident in the greater university area in the neighborhood of Oakland. Oakland houses the University of Pittsburgh, Carnegie Mellon University (CMU), and several other institutions of higher education. Oakland in particular has a dearth of Black-owned businesses. In Figure \ref{fig:Black-owned business_map}, we highlight the university community in Oakland in relation to the location of Black-owned businesses. In light of this, in this work we aim to support Black-owned businesses in Pittsburgh by focusing on the lack of market penetration in Oakland and specifically the university community.

In this work we describe the motivation, design, and implementation of an online platform, 412Connect (\url{https://www.412connect.org/}), launching in the Fall of 2021. The platform's purpose is to bridge the divide between university students and Black-owned businesses. Specifically, our target populations are a subset of Pittsburgh Black-owned businesses curated by our community partners and students at the University of Pittsburgh (with a focus on incoming students at the undergraduate and graduate levels). To better understand the challenges facing our target populations, we conducted surveys in the Winter of 2020/2021. Synthesizing these results following discussions with our community partners (Black Action Society, Blackowned.pgh, Cocoapreneur, Community Forge, the Pittsburgh Hub of Data for Black Lives, GSPIA Students of Color Alliance, Riverside Center for Innovation, Vibrant Pittsburgh, Western Pennsylvania Regional Data Center), we began work on the 412Connect platform. This platform seeks to promote and increase the visibility of Black-owned businesses in Pittsburgh and to expand their business presence on the University of Pittsburgh campus and in the Pittsburgh community. Specifically, 412Connect serves the needs of Black-owned businesses by highlighting relevant business information (websites, social media pages such as Twitter, Instagram, etc.) and matching these businesses to students with compatible interests. For participating students, 412Connect simplifies the process of discovery of Black-owned businesses in Pittsburgh and encourages continued engagement with the platform through the use of non-monetary incentives (badges, university recognition).

A summary of our key contributions and findings is as follows:

\begin{itemize}
    \item We identify two relatively unlinked populations with mutually compatible incentives. Building on previous attempts at connection that have often lacked impact or resulted in unintended consequences, we propose a novel model for the linking of university students to local Black-owned businesses (Section \ref{sec:existing_approaches}). Our approach leverages the unique access we have as members of the academic community, and may be relevant in other university areas juxtaposed with a diverse metropolitan community.
    \item In coordination with our community partners, we designed and implemented our online platform so that it may serve as a useful example for other socially conscious platform designers. Specifically, in our implementation we address the safety and integrity of Black-owned businesses, the reputation of our community partners, and the advantages and limitations of energizing a student population.
    \item We propose simple, novel models for badge design and content recommendation that are tailored to our application. For the problem of badge design, we prove a number of structural properties that crystallize our intuition for these systems and guide our implementation (Section \ref{sec:badge_design}). For content recommendation, we propose a natural linear programming formulation that balances maximizing the number of matches between students and businesses against equitably spreading traffic to all businesses (Section \ref{sec:recommendation_system}).
\end{itemize}

\subsection{Literature Review}\label{sec:lit}

Our work intersects with and draws from several streams of empirical and theoretical literature. Here, we overview some of these streams and explain how our work contributes to and/or differs from each. 

\subsubsection{Community-Based Platform Design for Social Good}

Our work contributes to a body of literature documenting and implementing online and/or community-based platforms that offer algorithmic solutions to multi-faceted societal problems. Academic collaborations leading to publicly implemented platforms have been deployed in a wide number of problem specific domains, we will highlight three here.  \citet{wilder2018equilibrium, wilder2020clinical} build a site which identifies influential nodes in partially-known networks for the dissemination of public health information. \citet{shah2017spliddit,flanigan2020neutralizing,flaniganfair} implement a website to help determine fair and transparent allocations. \citet{rubinstein2020dynamic,shi2021recommender} create platforms for the coordination of food rescue. To the best of our knowledge, our platform is the first academic approach to connecting Black-owned businesses with student markets.

\subsubsection{User Incentivization and Badge Design}

Our site recognizes user participation via \textit{badges}. A \textit{badge} is a non-financial award that is commonly given in recognition of user effort, especially in online settings. \citet{facey2017educational} survey some of the empirical and theoretical literature on online badge systems. The purpose of badges is many-fold. \citet{wilson2012review} perform a social psychology taxonomy of online badges, noting they are useful for goal-setting, motivating and affirming users, and as status/reputation symbols. Similarly, the motivations for volunteer populations on online platforms are diverse \citep{holdsworth2010volunteer}, and thus the interactions between users and badges is complex. \citet{burke2009feed} find that for new users who are inclined to contribute, receiving positive feedback (e.g. a badge) is a strong predictor of increased engagement. Further, there exists well-developed literature studying how users behave in online platforms \citep{wilson2012review,naaman2010really}, and what motivates them \citep{burke2011plugged,frey201817}. One key finding from these works is that users, especially volunteers, who engage with features like badges perform better on their goals \citep{burtch2021peer}. These works motivate our use of badge mechanisms for our student volunteer users.

On the theoretical side, an emerging stream of papers in computer science, economics, and operations research have attempted to explicitly model and analyze badge mechanisms. \citet{immorlica2015social} look at the design of badge mechanisms from a mechanism design perspective, treating badges as a form of social currency. \citet{anderson2013steering} model badges as attracting users to perform additional work across multiple dimensions of contribution. In a similar vein, \citet{chen2020motivating} investigate the efficacy of targeting and public/private recognition on eliciting prosocial behavior on Wikipedia. \citet{gallus2017fostering} also looks at soliciting reviews on Wikipedia by using symbolic badges and finds them effective. In our work, we provide a model of badges based primarily around user motivation, abstracting away reputational effects due to user anonymity concerns and simplifying the multidimensional nature of the badges studied in \citet{anderson2013steering}. While simple, our model allows for crisp structural characterizations of the optimal badge design model. We hope to validate and revise our model when the site goes live, incorporating observations of how users behaviour actually changes when exposed to badge systems.

\subsubsection{Content Recommendation Systems}

Our site also acts as recommendation system for users looking to discover local Black-owned businesses. Recommendation systems are an important and well-studied problem in industry that are driving a huge amount of academic attention over the last decade \citep{melville2010recommender,2020Verge}. Recent work on recommendation systems has employed techniques from the bandit literature, adapting the canonical explore and exploit frameworks to create recommendation systems that are explainable, equitable, and maximize user satisfaction \citep{mcinerney2018explore,mehrotra2019jointly}. As our site is launching data-blind, many of these more sophisticated techniques are out of reach for our platform. Further, as a volunteer organization with limited development resources, approaches with simple implementations are preferable, and as such, we take a pared-down approach based on linear programming that is similar to approaches for assortment optimization for two-sided markets \citep{ashlagi2019assortment}. The approach most similar to ours is \citet{shi2021recommender}, who also make targeted recommendations using a linear program with a diversity constraint to ensure that content is equitably spread across participants in the context of food rescue.

The remainder of this paper is organized as follows. In Section \ref{sec:motivation}, we dive into the challenges facing Black-owned businesses before and during the pandemic, present our survey results, and synthesize these into design principles for the 412Connect platform. In Section \ref{sec:implementation} we describe the 412Connect platform implementation and recruitment, placing particular emphasis on societal and ethical considerations. Finally, in Section \ref{sec:conclusions}, we conclude and highlight avenues for future work.

\section{Platform Motivation and Market Surveys}\label{sec:motivation}
In this section, we address specific socioeconomic challenges experienced by Black-owned businesses both in the United States and in Pittsburgh specifically, emphasizing the harmful consequences of the COVID-19 pandemic on the stability of Black-owned businesses. Next, we discuss existing approaches to enhancing visibility for Black-owned businesses and the successes and setbacks of these approaches which have informed the design of our platform. Finally, we discuss relevant results from our surveys of local Black-owned businesses and University of Pittsburgh students which lead into our platform design principles. 

\subsection{Challenges for Black-Owned Businesses} %

Black-owned businesses face specific challenges, some of which are outlined here. As of 2018, Black-owned employer businesses made up 2.2\% of all American employer businesses \citep{House2018}. Although these businesses have proven to be pillars of their communities by creating jobs and hiring locally, Black-owned businesses face persistent levels of systemic discrimination, inequality, ostracism, and more. Additionally, Black-owned businesses on the whole lack access to mentorship networks, business opportunities, and capital through traditional financial institutions \citep{House2018}, and are more likely to report financial distress than white-owned businesses \citep{Misera2020}. Consequently, Black-owned businesses tend to have lower revenue than white-owned businesses, with 45\% of Black-owned businesses reporting less than \$100,000 annual revenue as compared to only 18\% of white-owned businesses \citep{Mills2020}. This reveals the importance of supporting Black-owned businesses as a step toward enhanced economic growth, economic equality, and closing the racial wealth gap \citep{House2018, Misera2020}. 

\subsubsection{Black-Owned Businesses in the COVID-19 Pandemic}

While the pandemic has introduced challenges for all business owners, the economic impacts have disproportionately affected Black-owned businesses. Factors like lockdown procedures, declines in demand, and lack of access to equitable Paycheck Protection Program (PPP) loan coverage, as well as pre-existing gaps like insufficient cash funds and bank relationships, have led Black-owned businesses to suffer more closures and declines in available funds than any other racial group. Since the beginning of the pandemic, Black-owned businesses have closed at a rate of more than two times that of their white counterparts, and cash balances have decreased more than nine times that of their white counterparts \citep{Misera2020}. Between February and April of 2020, the number of working Black business owners decreased by a staggering 41\% in comparison to 32\% Latinx owners, 26\% Asian-American/Pacific Islander owners, and 17\% white owners \citep{Mills2020}. Geographic location plays a significant part in this alarming distribution, with Black-owned businesses largely located in metropolitan areas that have been more severely affected by COVID-19 and the resulting lockdowns. In fact, two-thirds of American counties with the highest population of Black-owned businesses are in the top fifty counties most severely affected by COVID-19 \citep{Mills2020}. The severity of the economic struggles of Black-owned businesses during the pandemic combined with the increased visibility of police violence against Black Americans has led to an effort to support Black entrepreneurs and their businesses. These factors played a large part in motivating the conception of both our platform and other similar platforms. In the next subsection, we reflect on these similarly motivated implementations and contrast them with our work.

\subsection{Existing Approaches}\label{sec:existing_approaches}

There have been many attempts to support minority-owned businesses in the United States. Some of these attempts originated from within the minority group, while others were initiated by individuals or entities outside of the minority group. In Pittsburgh, there are several Black organizations that are dedicated to incubating, elevating, and/or promoting Black-owned businesses. Efforts range from offering financial and managerial training to compiling lists of businesses and promoting them through a website or social media account. Large corporations in the technology sector like Google and Yelp have also added Black-owned business search functionality to their websites \citep{verge_yelp,verge_google}. 

Each of these approaches have their own strengths and weaknesses. Organizations led by and operating within a minority group are more able to foster a relationship of trust and connect more deeply with Black-owned businesses. This can lead to a more open exchange of information about a business's specific preferences and constraints, enabling ``insider'' organizations to provide impactful support. These ``insiders'' may, however, be less connected with the broader market, especially when markets are geographically segregated e.g., as is the case in Pittsburgh. 

Organizations from outside the community may have the opposite experience, in that these ``outsiders'' may not understand the Black-owned businesses very well, but may have access to a wider, more general, market. However, this lack of insider information can quickly lead to negative externalities for Black-owned businesses. For example, in the case of one company promoted in Target ads and subsequently bombarded overnight with negative reviews, we learned that simply promoting Black-owned businesses to a general audience may attract people who intend on harming Black-owned businesses \citep{cnn_target}. 

For both insiders and outsiders, there are advantages and disadvantages that may mitigate their impact. In light of this, one motivation for the development of our platform is our ability to walk the line between these extremes and offer effective support to Black-owned businesses in conjunction with our community partners while reaching a larger mainstream market governed by norms that would discourage users from abusing the platform. It is on these principles that we built 412Connect. 

\subsection{Market Research}\label{sec:surveys}

Having identified our two target communities, Black-owned businesses in the greater Pittsburgh area and university students in the neighborhood of Oakland, we conducted multiple surveys to better understand these populations and the barriers between them.

\subsubsection{Student Survey}\label{sec:student_survey}

 412Connect's Community Engagement Team performed an exploratory survey in March of 2021. They administered a 5-minute survey online to 24 members of the University of Pittsburgh student body. This survey found that 92\% of respondents have a desire to support and patron Black-owned businesses but only 50\% could name a local Black-owned business. Further, respondents expressed that their greatest barriers to interaction with Black-owned business were a lack of knowledge about businesses and their services (92\%), financial constraints (42\%), and lack of transportation (21\%). The survey also showed that students (considering the aforementioned constraints) are most likely to support Black-owned businesses through social media interaction and that they are most likely to visit Black-owned businesses in the food industry. Additionally, participants stated that they are more likely to visit businesses with good “word of mouth” reputation (100\%), a strong online presence (92\%), and unique services (83\%).

\subsubsection{Business Survey}\label{sec:business_survey} 

In collaboration with 412Connect's Community Engagement Team, Community Forge and the Pittsburgh hub of Data for Black Lives surveyed 22 Pittsburgh Black-owned businesses in February and March of 2021. They found that 85\% of respondents ranked reaching new customer bases as an extremely important goal, with 61\% feeling inadequately linked to their community. Specifically, we saw a strong desire to increase connection to local students, with 85\% feeling inadequately linked to the university community and wanting to improve this connection, while 54\% desired increased social media support.

\subsection{Platform Proposal}\label{sec:proposal}

Based on the need uncovered in our surveys, and in partnership with various community organizations (including nonprofit organizations, incubators, and businesses) in the Pittsburgh community, we have designed and created the 412Connect platform to encourage students to engage with Black-owned businesses via online interaction. Since, as opposed to purchasing goods or services, social media (likes/posts) and search engine activity with businesses are relatively inexpensive and accessible actions for most students, we aim to use our platform to facilitate these online interactions. Specifically, in Section \ref{sec:implementation} we describe how these insights directly influence our implementation decisions through the conception of the platform.

\section{Implementation of the 412Connect Platform}
\label{sec:implementation}
In this section, we first overview the platform's implementation and operation. Next, we discuss our platform's engagement and recruitment strategies, as well as the important organs of the platform, including the badge system and dashboard. Finally, we discuss platform deliverables like volunteer credit and statistical engagement reports, and conclude with ethical considerations. 

\subsection{Overview of 412Connect Platform}

\begin{figure}[ht]
    \centering
    \includegraphics[width=0.95\linewidth]{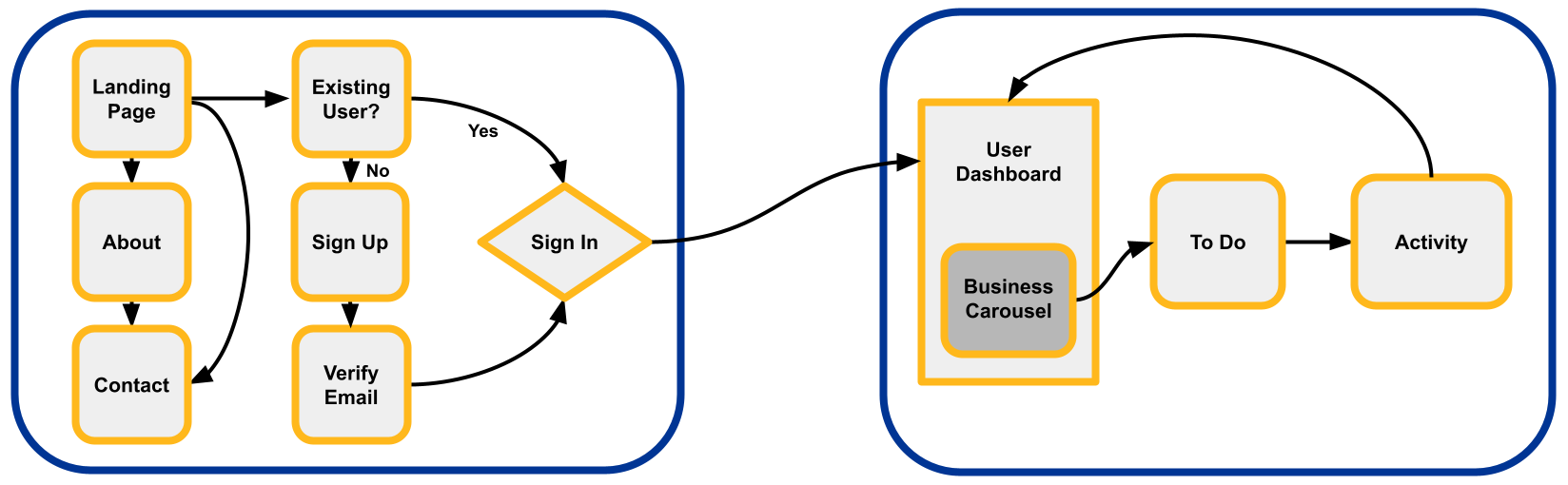}
    \caption{Operational Flowchart of User Website Interaction {\rm The flowchart explains the possible choices and actions a user can take, starting from the initial landing page to sign-up and through activity completion.}}
    \label{fig:flowchart}
    \Description[Once users sign in, they can browse businesses and are guided to specific tasks for each business.]{To create their account, users must verify their email. After sign in, users are presented with a dashboard which includes their current badges and recommended businesses. Clicking on any one of this businesses presents them with the business page including links to social media and tasks to be completed.}
\end{figure}

We have built the 412Connect platform in collaboration with university and community volunteers (\url{https://www.412connect.org/}). The site displays participating Black-owned businesses along with business-specific activities users can complete. The platform further operates to incentivize members to interact with these businesses, either to \textit{Explore}, where the user is asked to engage with prepared business content, or through \textit{Social Media}, where the user can follow or like the business on Twitter, Instagram, and/or Facebook (Figure \ref{fig:wire_business}). Users are able to track their engagement with Black-owned businesses through a dashboard on the 412Connect platform (Figure \ref{fig:wire_dashboard}). At regular intervals, businesses will be given a report detailing aggregate engagement statistics, such as new customers generated, new followers on social media, and more. An overview of the site's functionality from the users perspective is presented in Figure \ref{fig:flowchart}. Next, we describe our strategies to attract businesses and users to our platform.

\subsection{Recruiting Platform Participants}

In this section, we discuss engagement and recruitment strategies for local Black-owned businesses, community partners, and student users. 

\subsubsection{Black-Owned Business Curation}\label{sec:curation}

Utilizing principles from Community-Based Participatory Research (CBPR) \citep{Hacker2013}, we have worked to foster strong relationships with the Black entrepreneurial community. Building on these relationships, we conducted interviews and surveys with Black-owned businesses to understand the issues they face and their potential motivations for joining our platform. However, community partners have expressed trepidation about working with university groups, as university projects have often stepped on the toes of existing community institutions. Moreover, there is an existing ecosystem of community organizations working to support Black-owned businesses in Pittsburgh, some of which we have entered into conversation with, notably: Cocoapreneur Pgh, Community Forge, Blackowned.pgh, Vibrant Pittsburgh, Black Action Society, the Pittsburgh Hub of Data for Black Lives, Western Pennsylvania Regional Data Center, Riverside Center for Innovation, and the GSPIA Students of Color Alliance.

In light of this tension, instead of directly recruiting businesses to our platform, we implement a \textit{curation model}. Our \textit{curation model} is a model of recruitment where a community member with extensive ties to Black-owned businesses in Pittsburgh is invited to act as the curator. The curator selects a set of Black-owned businesses who will participate on the platform for a fixed period of time (on our platform, one month). The curator assists in gathering information from each participating business so that their engagement with our platform can be as beneficial as possible. This information includes: business growth priorities (i.e. social media interaction, consumer education), growth objectives (i.e. XXX new Instagram followers, XXX people learning about a business), and desired date of completion. Partners acting as curators will be credited as collaborators in associated events and services as well as given access to directories and partners. 

\subsubsection{Student Recruitment}\label{sec:student_recruitment}

Building off our motivations outlined in Section \ref{sec:motivation}, the user base of our platform will be Pittsburgh university students of all affiliations (undergraduate, graduate, professional), contained primarily in the neighborhood of Oakland. First, we will advertise broadly to the university community via departmental mailing lists, and via contacting established socially conscious student groups like Student Government, the Diversity and Inclusion Committee, GSPIA administration, and the GSPIA Students of Color Alliance. Second, we will put particular emphasis on recruiting incoming students. Incoming students are eager to create personal and professional networks at their new school and, for many, their new home in Pittsburgh. To capitalize on this recruitment opportunity, the platform's launch will coincide with the start of the academic year. Finally, we are coordinating with the university administration, specifically the Office of Student Affairs, to allow student users to gain volunteer credit through Outside the Classroom Curriculum (OCC) by interacting with our platform. 

\subsection{Platform Design and Features}

In this section, we will describe our platform's features in detail. These features include two modes of engagement with businesses, a badge system to track and reward user progress, and a content recommendation system to refer students to  businesses based on their reported preferences. 

\subsubsection{Modes of Engagement: Explore and Social Media}\label{sec:modes_of_engagement}

On the 412Connect platform, there are two categories of engagement with Black-owned businesses which we term \textit{explore} and \textit{social media}. An \textit{explore} activity involves answering curated business questions about media that the business gives to the site (i.e. articles, promotions, unique services, etc.). A \textit{social media} activity involves interacting with businesses' social media accounts (Facebook, Instagram, Twitter). Businesses will work with the curator and 412Connect to create activities as described in Section \ref{sec:curation}.

\subsubsection{Badges} \label{sec:badge_description}

In order to help users track progress and incentivize participation, we have designed a badge system. Badges exist for both modes of engagement. Each social media or explore activity counts towards thresholds for three badge tiers (bronze, silver, gold). Specifically, when the user completes their first activity, they will be awarded a bronze badge; when the user completes three activities, they will be awarded a silver badge; and when the user completes six activities, they will be awarded a gold badge. These threshold levels (1, 3, and 6) are derived from a mathematical model, the details of which are provided in Section \ref{sec:badge_design}. The badges can be seen in Figure \ref{fig:badges} in the Appendix.

\subsubsection{Business Recommendations} \label{sec:business_carousel}

Businesses displayed on the site fall into one of four categories: beauty, entertainment, food, and shopping. Businesses are displayed to users one at a time via a carousel shown in Figure \ref{fig:wire_dashboard}. We allow users to articulate preferences over business types and specific services in order to receive a tailored set of business recommendations. We perform this customization and recommendation by solving a particular linear program for each curation period, the details of which are provided in Section \ref{sec:recommendation_system}. The output of the recommendation system is a set of probabilities, unique to each user, which represent the probability of a business being displayed first. Every time a user visits their dashboard, the order of the businesses in the carousel will be randomly drawn with weights equal to the recommended probabilities. The display profile for each business contains basic information about the business, its location, and its explore and social media activities. An example of a business profile is shown in Figure \ref{fig:wire_business} in the Appendix. 

\subsubsection{Student Dashboard}

Each student user of the 412Connect platform has their own dashboard, displayed to them when they log in. It is composed of their personalized business carousel (Section \ref{sec:business_carousel}) and a display of the user's complete badges and progress toward incomplete badges. An example of a student user's dashboard with their badge progress is shown in Figure \ref{fig:wire_dashboard} in the Appendix. 

\subsection{Deliverables}

In this section, we describe the platform's outputs over each curation period. Specifically, these are impact reports to our stakeholders, which will have statistics to help quantify the effect of the 412Connect platform.

\subsubsection{Curator Reports}

Curator reports will be composed of their own organization report along with the reports for all of the businesses. They will also be provided with the results of a student participant survey.

\subsubsection{Black-Owned Business Reports}

Business reports will summarize engagement received over the period, both on 412Connect and on external social platforms.

\subsubsection{Academic Department Reports}

Academic department reports will include both results from the student survey and breakdowns of student engagement across academic departments.

\subsubsection{Student Reports}

Student reports will consist of an individual breakdown of engagement, a group report on the effect of 412Connect, and whether or not the student was awarded OCC credit, as mentioned in Section \ref{sec:student_recruitment}.

\subsection{Ethical Considerations and Oversight}

In building the 412Connect platform with our community stakeholders, one of the things that became clear was their disillusionment with previous university partnerships. Things that were of notable concern include the ownership of data, following through on promised features, and the overshadowing of organizations that had already been working in the community for many years. This has informed our planning, software development processes, and promised features at launch.

To mitigate some of these potential issues, businesses will enter our platform after a discussion which includes our curator and our team so as to communicate potential concerns, explain the platforms goals, and increase participant enthusiasm. Further, in order to foster oversight on our team by those with perspectives and backgrounds different than ours, the project is forming a Community Advisory Board. It is composed of knowledgeable community members with special interest in racial equality to inform sustained efforts and collaborative action. 

A critique of our platform could be that it does not directly encourage monetary engagement with businesses. We limited our focus to online interactions because, as shown in Section \ref{sec:surveys}, businesses have a genuine interest in growing their online presence, students are very willing to connect with businesses online, and creating awareness is a necessary prerequisite to monetary transactions. Finally, to address the concerns of negative online feedback outlined in Section \ref{sec:existing_approaches}, we place all participating businesses behind the login screen which only verified university members may access. This also provides us with a disciplinary structure for enforcement if users are found to be using the site to spread hate.

\section{Conclusion}\label{sec:conclusions}

In this work, we detailed the challenges of Black-owned businesses in the United States and city of Pittsburgh. Taking into account local dynamics and the communicated desires of Black-owned businesses in the Pittsburgh region, we determined that university students represent an under-utilized market for these businesses. We investigated the root causes for this inefficiency. In our survey results, students reported that they are limited by low finances, inaccessible transportation, and lack of knowledge about local Black-owned businesses. Therefore, we have created a platform to allow university students to engage with and support Black-owned businesses.

Through business curation, recruitment events, non-monetary incentives, and progress-tracking badge systems, we engage our student users while directly supporting individual businesses through engaged listening. This work is made possible by a network of community partners, including nonprofit organizations, incubators, businesses, individuals, and university administrators. The long-term aim of project is to create an inclusive network of support for Black-owned businesses, and potentially play one small part in closing the racial wealth gap in Pittsburgh.

\subsection{Potential for Future Developments}

We believe our work has the potential for significant iteration and improvement. When the platform goes live in August 2021, we will be able to collect data on our site users and how they interact with the platform. We hope to use that data to validate our badge model and to enrich our business recommendations so as to maximize user engagement. On the theoretical side, we believe there is significant additional work to be done on badge mechanisms, specifically on their interplay with the recommendation system. We hope to show that the badge system and recommendation system have a symbiotic relationship, where badges implicitly select high engagement users, and that information can be fed into the business recommendation system to yield explainable, high quality, recommendations. Finally, we believe that the conditions that motivated this platform, namely, a diverse marginalized community surrounding a disconnected university environment, are not unique to Pittsburgh. If successful in Pittsburgh, we will explore broader deployment at other institutions and in other cities.

\begin{acks}
The 412Connect platform is the result of a collective effort of more than 50 volunteers and partners associated with the Center for Analytical Approaches to Social Innovation (CAASI) Grief to Action initiative. We want to especially acknowledge the project management team (I Younan An, Mara McCloud, Ivy Bryony Chang, Tyler Olin, Claire Guth, and Collin Griffin) and our scavenger hunt curator (Adam Gerard and Juan Garrett from Riverside Center For Innovation). The following people shared/collected the BOB data, conducted surveys, and/or managed community outreach: Kyley Coleman (Blackowned.pgh), Khamil Scantling (Cocoapreneur), Alex Jackson (Pittsburgh Hub of Data for Black Lives), Herman Johnson (Community Forge), Miguelina Javier, Diane Roth Cohen, Sonya Akhgar, Sarah Downing, Chloe Harvey, Caelan Schick, Vannie Wang, and Falon Weidman. We thank our volunteer tech team: Jeremy Olin, Andi Halim, Gurmail Mathon, Kevin DeMaoiribus, and Melanie Cooray. We also thank Angel Christou and Kayla Pierre (Black Action Society), Minnie Jones (Vibrant Pittsburgh), Uchenna Mbawuike (GSPIA Students of Color Alliance), David Walker and Bob Gradeck (Western Pennsylvania Regional Data Center) for their support. We thank and Rob Racuna from Pitt Innovation Office and Jennifer Tseng from University Counsel for their legal guidance. Finally, we thank the Office of Community Engagement at the University of Pittsburgh for their financial support through the ESDI grant and GSPIA for CAASI startup funding.
\end{acks}

\bibliographystyle{ACM-Reference-Format}
\bibliography{p412}

\appendix

\section{Theoretical Components: Model and Analysis for Badge Design and Business Recommendations} \label{sec:analysis}

In this section, we build on the insights derived from our surveys described in Section \ref{sec:surveys} to propose models that guide our understanding and implementation of our badge system (Section \ref{sec:badge_description}) and business recommendation system (Section \ref{sec:business_carousel}). For badge systems we introduce a new model based on renewing user motivation and prove structural characteristics of optimal badge systems. We then decide our preliminary badge thresholds by fitting the model to our parameters. For business recommendation, we introduce a simple and equity-oriented approach based on linear programming (LP). 

\subsection{Badge Design}\label{sec:badge_design}

In this section, we will propose a new badge model based around user motivation, a feature of badges highlighted in \citep{facey2017educational}. Specifically, we imagine badges as a mechanism to mitigate user fatigue by \textit{refreshing} the user's motivation to continue participating in the platform, and propose a simple model to capture this depiction of badges. Under this model, we prove several intuitive structural results about the design and implementation of these badges. Finally, we solve our model and propose initial badge thresholds for the launch of our platform.

\subsubsection{Model and Motivation}

Recall on our platform we use badges to reward users for completing two types of engagement activities: explore, and social media, as discussed in Section \ref{sec:modes_of_engagement}. In our system the badges are tiered according to the level of recognition the badge represents, thus for each badge type there are a number of thresholds for the various levels of effort required to earn that tier of the badge (i.e. bronze, silver, and gold). In Figure \ref{fig:badges}, the three tiers of badges for both activity types are shown. We refer to the specific number of activities necessary to achieve a tier of the badge as the \textit{threshold} for that badge tier. We are interested in models that help us decide where to place these thresholds so as to optimally induce and recognize platform participation. 

To study badges for this purpose, we model users as having idiosyncratic motivations to complete activities for a badge, where that motivation is modeled as a number drawn independently from a known prior distribution $F$\footnote{Note our user-base is has complex and multifaceted motivations \citep{holdsworth2010volunteer}, in our model these varying motivations are all folded into the prior distribution.}. We further assume the number of activities a user completes is independent of the businesses themselves and depends only on their intrinsic motivation. The badges themselves act as an intervention which \textit{refreshes} the users motivation. Specifically, when a user completes enough activities to cross the threshold and earn a badge, we model the users renewed motivation as a fresh draw from their motivation distribution $F$.

We now formally describe our model and its mechanics. For a badge with multiple tiers\footnote{Exogenously determined, the number of tiers is three in our platform.} we model the platform as deciding $K \in \mathbb{N}$ thresholds for the allocation of badges $\{T_i\}_1^K \in \mathbb{R}^K$, where $T_i$ is the work required to earn the $i^{th}$ badge tier given the user has already earned the $(i-1)^{th}$ badge tier. Once thresholds have been set, users interact with the system by drawing a number of activities from their motivation distribution, $V_1 \sim F$. If $V_1 \ge T_1$, the user will complete $T_1$ activities, at which time they are awarded the first tier of badge. The users motivation is then refreshed, i.e., they draw a second time from their type distribution, $V_2 \sim F$. If $V_2 > T_2$, then the user performs $T_2$ activities, gets a second tier of the badge, and the goes for the third, and so on. The user leaves the system (stops participating) the first time when $V_i  < T_{i}$, if they obtain the final badge tier we assume the users redraws once more\footnote{We note this is equivalent to including a phantom final badge for which the additional work required to obtain it is infinite.}. Under this model, the objective is to find thresholds $\{T_i\}_1^K$ that maximize the expected number of activities completed by a user.

\subsubsection{Structural Analysis of Optimal Badge Policies}\label{sec:optimal_badge_policy}

In this subsection, we will demonstrate a number of attractive and intuitive properties about the structure of optimal threshold policies. These results will guide our implementation and may be useful for other platform designers seeking to implement badge based designs. Before we state the results of this section, we will first introduce some useful notation. Let $G_1(t)$ be the maximum expected number of activities completed by a user with motivation drawn from $F$, facing a badge with one tier awarded at threshold $t$, and let $\cF(x) = 1 - F(x)$. Under our model, if $V_1 \ge t$, then the expected number of activities is $t + \E[V]$, since they will perform $t$ activities, then redraw and perform $\E[V]$ more activities. On the other hand if $V_1 < t$, then the expected number of activities is $\E[V \I{V < T}]$ where $\I\{\cdot\}$ is indicator function. Thus can write out $G_1(t)$ as,
\begin{align}
   G_1(t)  &= \left(t + \E[V]\right) \cF(t) + \E[V \I\{V < t\}]. \label{eq:opt_one} 
\end{align}

When the optimal threshold is unique, we represent the optimal choice using $T_1 := \argmax_t G_1(t)$. For more than one tier level, we will let $G_i(t_1, \ldots, t_i)$  to be the expected number of activities performed when the first threshold requires $t_1$ completed activities, the second threshold requires $t_2$ \textit{additional} completed activities, and so on. Similarly, when unique, we will let $T_i$ be the optimal value for the $i^{th}$ threshold.

Finally, before we state our main results we will review one useful and common property of a distribution, log-concavity.

\begin{definition}[Log-Concavity]\label{def:logconcave}
A distribution $F$ with density $f$ is log-concave if $\log(f(x))$ is a concave function.
\end{definition}

For a very nice overview of log-concave distributions and their applications in a number of economic contexts see \citet{bagnoli2005log}. There are two especially important properties of log-concave distributions that make them relevant for modeling motivation distributions in our context. The first is that they have increasing hazard rates, which implies that for any user, instantaneously refreshing their motivation distribution increases the expected number of activities they will perform i.e. $\E[V | V \ge t] \le t+\E[V]$ for all $t$. The second property of log-concave distributions is that since their hazard rate is monotone, Equation \ref{eq:opt_one} has at most one critical point, we will prove this formally in Section \ref{sec:appendix_pfs}. Armed with these insights, we are now ready to state the main structural results of this section.

\begin{proposition}[Structure of Optimal Badge Policy]\label{prop:badge_work_inc}
Let $V \sim F$ be the distribution for user motivation. Then for an optimally chosen set of $K$ badge thresholds, the following properties hold:
\begin{enumerate}[label=(\alph*)]
\item If $F$ is log-concave, the optimal thresholds are unique and the additional work required to earn the next badge tier is monotone increasing i.e. for any $K$ and $i \in [K-1]$, $T_{i} \le T_{i+1}$.
\item If $F$ is log-concave, the maximum work required to earn a higher tier of badge is upper-bounded by a constant independent of $K$. Specifically if $h(\cdot)$ is the hazard-rate function of $F$, $T_i \le h^{-1}\left(\frac{1}{\E[V]}\right)$ for all $i$.
\item The marginal increase in expected induced effort provided by an additional level of badge is monotone decreasing i.e., $G_{i+1}(T_1,\ldots, T_{i+1}) - G_{i}(T_1, \ldots, T_i)$ is decreasing as $i$ increases. 
\end{enumerate}
\end{proposition}

Proposition \ref{prop:badge_work_inc} formalizes several intuitive properties about optimal thresholds for badge policies. Part (a) states that to earn higher tiers of badges, the amount of work required should also increase matching common implementations. However, part (b) states that the increase in work eventually tapers off to a maximum level independent of $K$, precluding, for instance, policies where the additional work required doubles at each tier. Finally, part (c) of the theorem states the benefit of additional badge tiers is decreasing, suggesting that most of the benefit is garnered by a small number of well-chosen badge tiers as opposed to many tiers.

\subsubsection{Preliminary Badge Threshold Computation}

\begin{figure}[ht]
\centering
\subfloat[][]{
    \label{fig:badges-social_bronze}
    \includegraphics[width=0.3\linewidth]{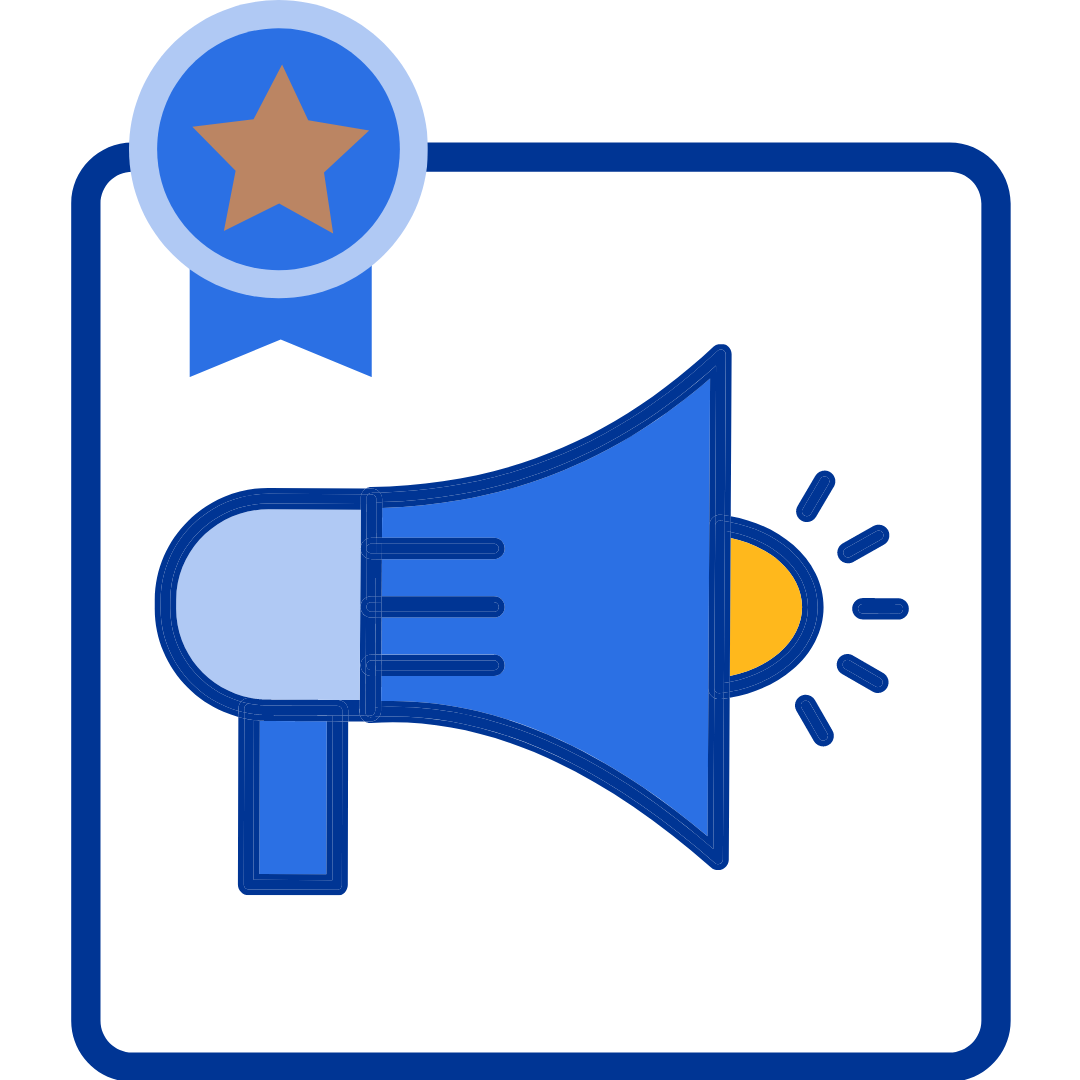}}
\hspace{8pt}
\subfloat[][]{
    \label{fig:badges-explore_bronze}
    \includegraphics[width=0.3\linewidth]{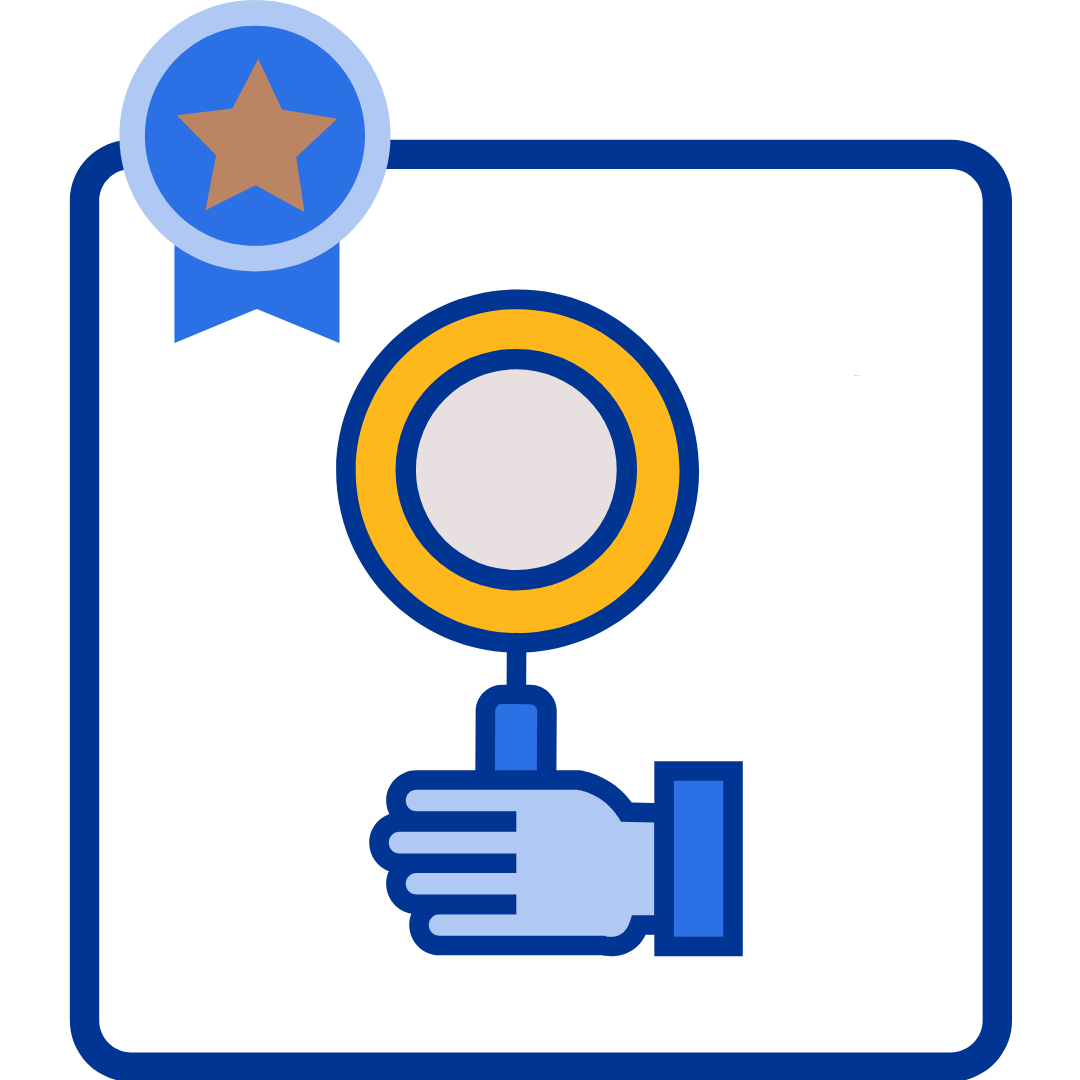}}
\\
\subfloat[][]{
    \label{fig:badges-social_silver}
    \includegraphics[width=0.3\linewidth]{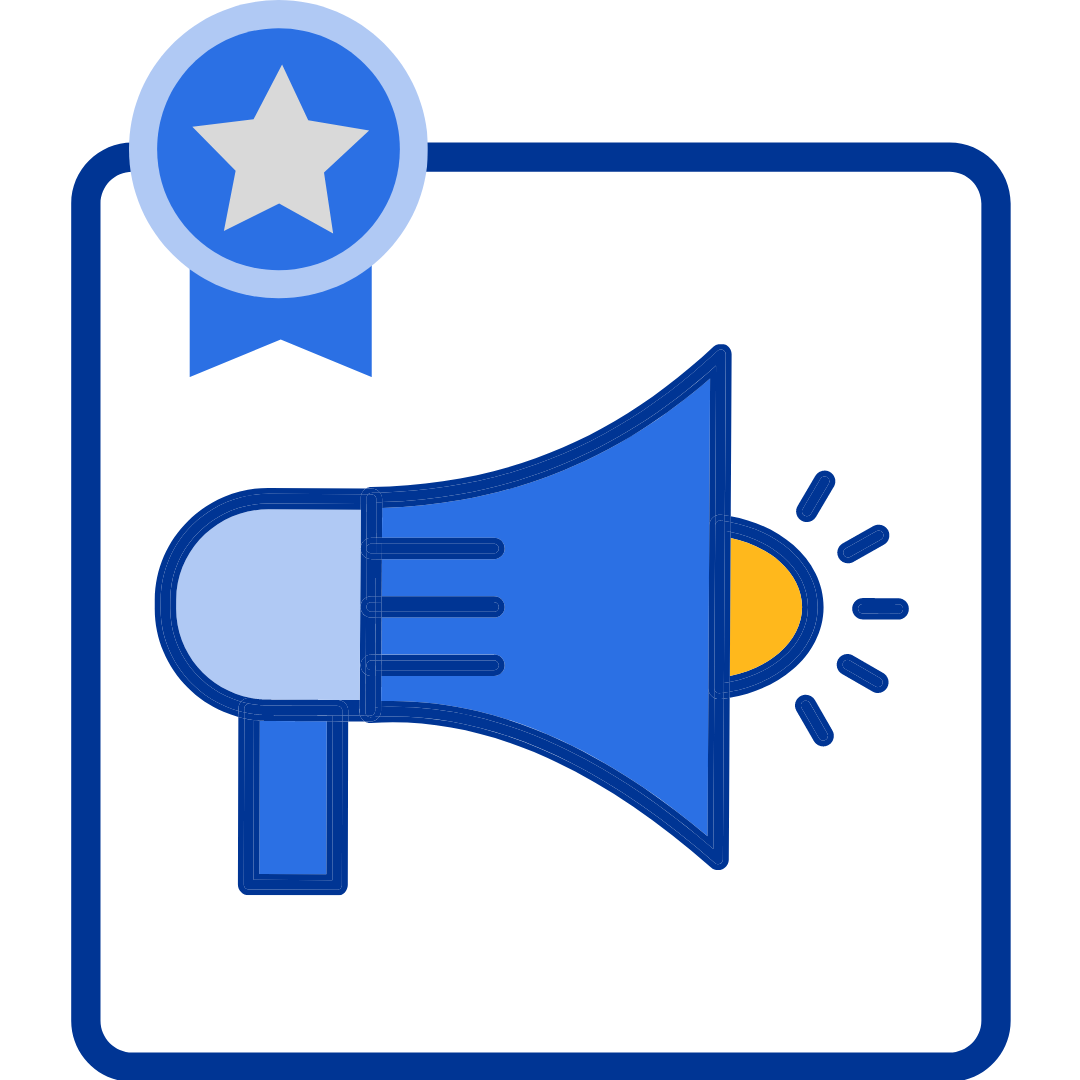}}
\hspace{8pt}
\subfloat[][]{
    \label{fig:badges-explore_silver}
    \includegraphics[width=0.3\linewidth]{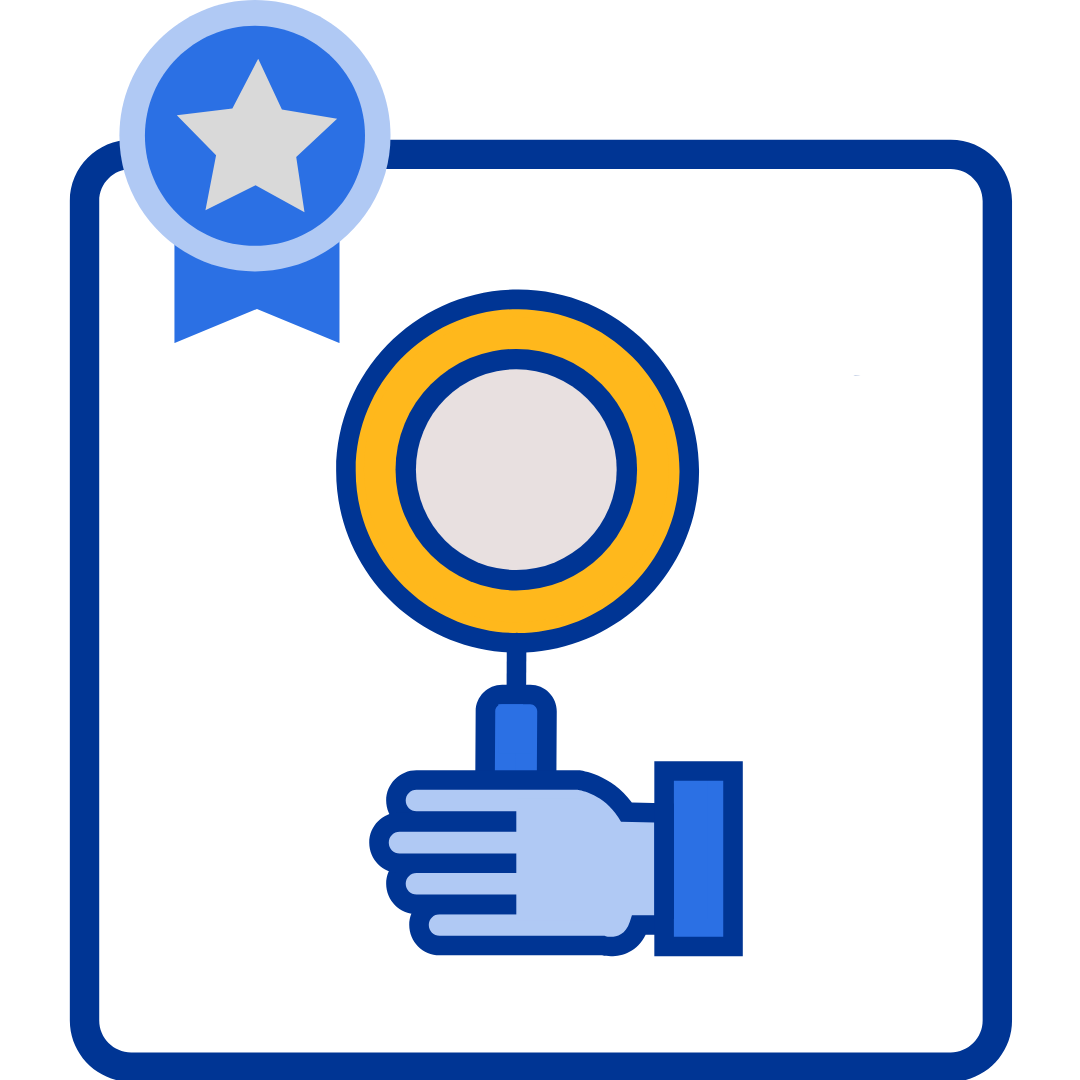}}
\\
\subfloat[][]{
    \label{fig:badges-social_gold}
    \includegraphics[width=0.3\linewidth]{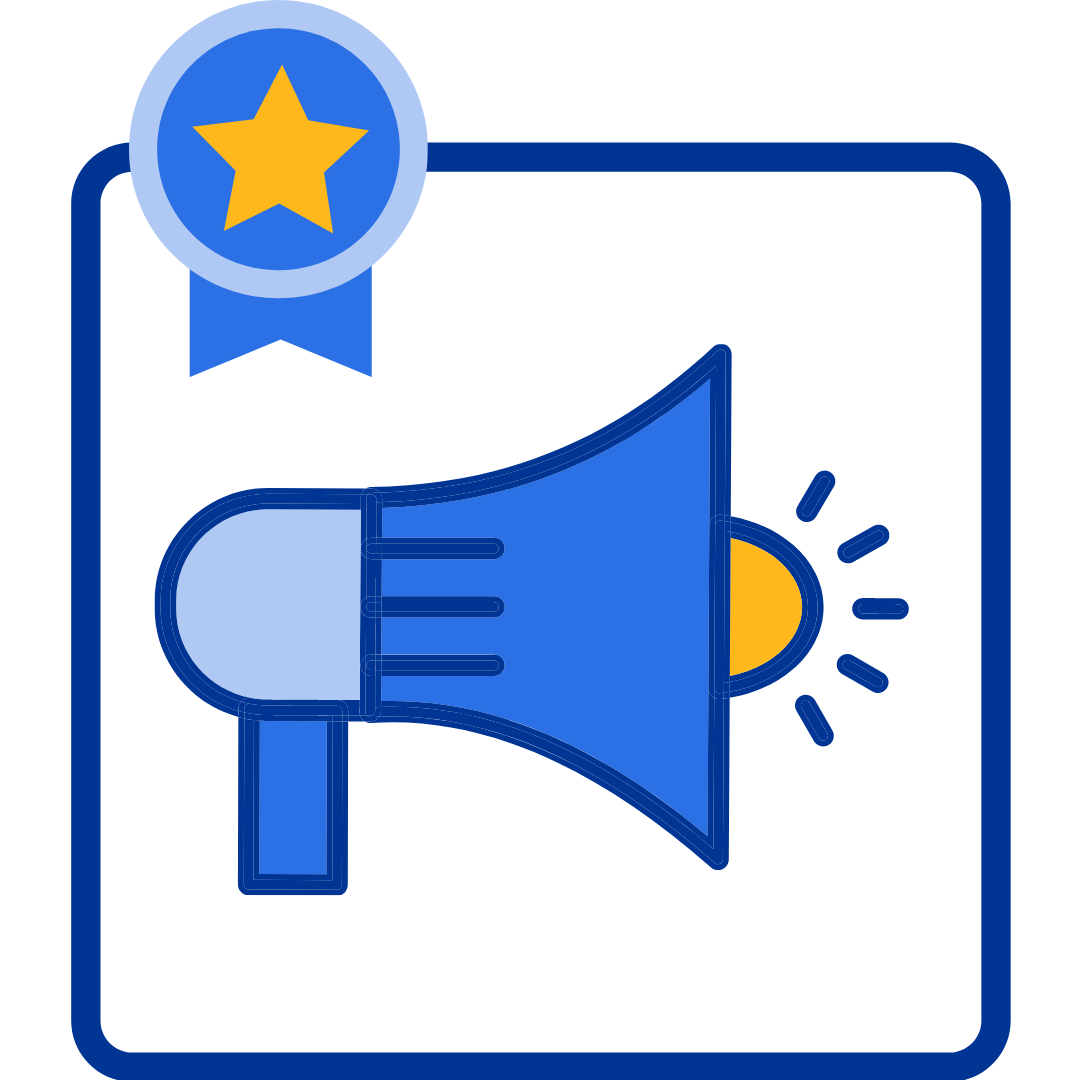}}
\hspace{8pt}
\subfloat[][]{
    \label{fig:badges-explore_gold}
    \includegraphics[width=0.3\linewidth]{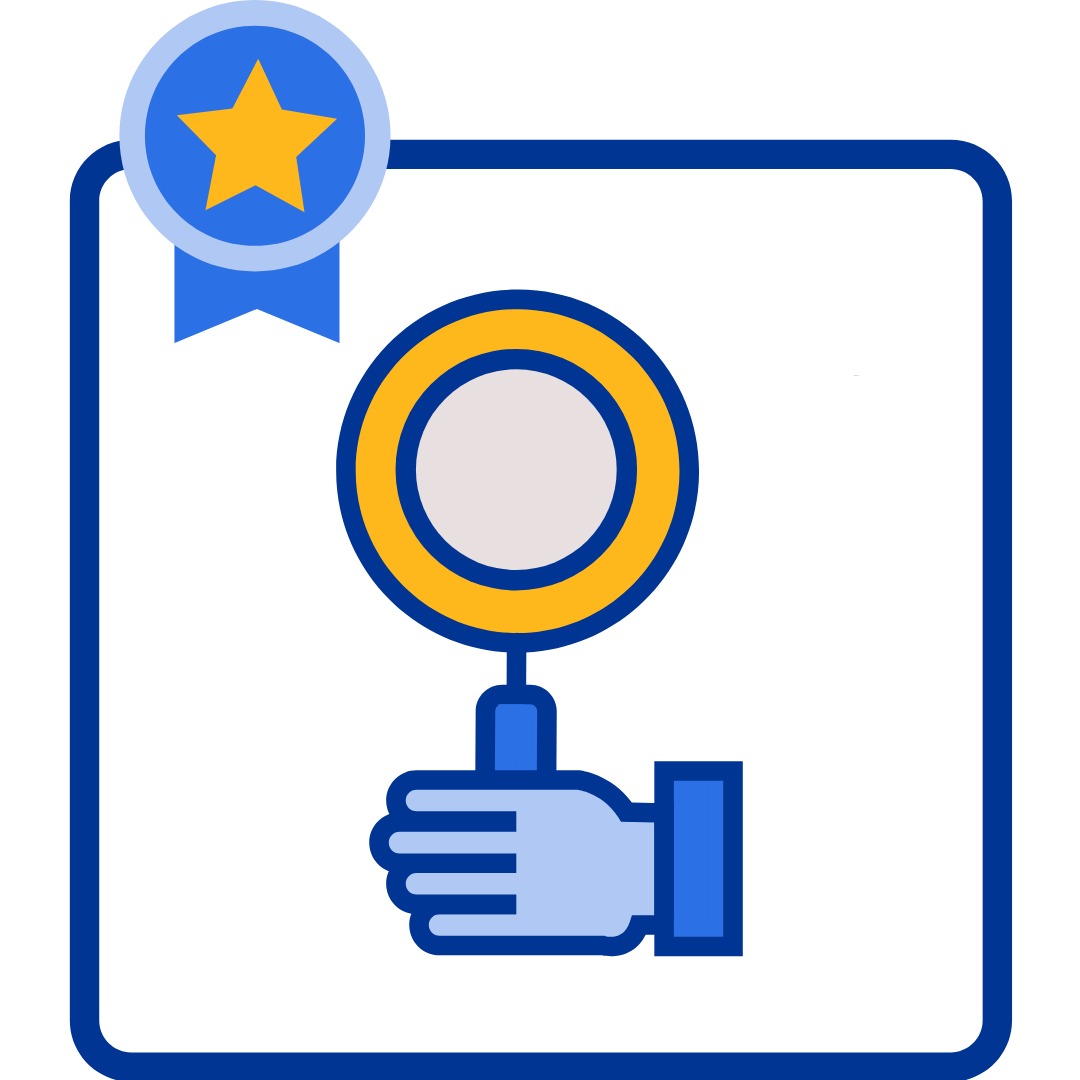}}

\caption{Tiered Badges for Social Media Engagement and Exploration. {\rm The set of badges on the left (\protect\subref{fig:badges-social_bronze}, \protect\subref{fig:badges-social_silver}, \protect\subref{fig:badges-social_gold}) are awarded for engaging with businesses on social media, while the set of badges on the right (\protect\subref{fig:badges-explore_bronze}, \protect\subref{fig:badges-explore_silver}, \protect\subref{fig:badges-explore_gold}) are awarded for learning more about participating businesses. Note that the badges are visually distinct (top to bottom: bronze, silver, and gold ribbons) with colors consistent with the additional effort required to earn them.}
}
\label{fig:badges}
\Description[User Badges]{Two categories of badges in three tiers (bronze, silver, gold) for users to collect. A megaphone represents Social Media Engagement and a hand holding a magnifying glass represents Exploration. Ribbons in the corner of each badge help to differentiate the tiers.}
\end{figure}

For the launch of our platform, we will solve an instance of our badge model using parameters estimated from our survey results, and expectations for the initial run of the platform. Specifically, for our first run we will have $K = 3$ tiers of badges (as shown in Figure \ref{fig:badges}) with a maximum number of activities a user can complete of less than fifteen since that is the number of businesses included in our first curation period. We will compute the optimal badge thresholds assuming the motivation distributions is Uniform[0,5] and Uniform[1,10], which represent pessimistic, and optimistic priors, over our target population, respectively. In the pessimistic case the (rounded) optimal thresholds are $T_1 = 1, T_2 = 2$, and $T_3 = 3$. In the optimistic case the (rounded) optimal thresholds are $T_1 = 2, T_2 = 4$, and $T_3 = 6$. For details see Section \ref{sec:badge_compute}. We opt for the pessimistic thresholds in our initial launch, but will revise after the initial run with thresholds fitted to the observed motivation distribution.

\subsection{Recommendation System}\label{sec:recommendation_system}

In this subsection we consider the problem of recommending businesses to users. We will model the problem as a generalized bipartite matching with fairness constraints, which ensures equity across the number of user-to-business matches. We then encode the model as a linear program (LP), the solution of which will be a set of user-specific distributions over businesses within the curation period, which we will leverage to make recommendations.

\subsubsection{Model and Motivation}

Our goal will be to find an assignment of businesses to users that maximizes the agreement between user's preferences over business types (service, food, and beauty) and activities that can completed (social and explore). Without any additional constraints, the problem can we written as a maximum weight bipartite matching problem where the weights are the expected number of activities a user will complete for each business. However, as we want every participating business to benefit from their participation on the platform, so we will further ensure that all businesses are given a similar number of impressions. Inspired by the $4/5^{th}$ rule used in law to determine disparate impact \citep{boardman1979another}, we will require that all businesses receive a number of impressions (i.e. show up in a users recommendations) such that the ratio between the businesses with the minimum and maximum number of impressions is greater than $4/5^{th}$ in expectation. Our objective will be to maximize expected number of activities that could be completed.

\subsubsection{Recommendation Linear Program} \label{sec:recommendation_linear_program}

We now formally describe our recommendation LP. To aid in our description we first set some notation.

\begin{itemize}
\item Let $b \in [B]$ be the chosen businesses in a curation period. 
\item Let $u \in [U]$ be the registered users at the beginning of curation period. 
\end{itemize}
The decision variables for the model will be $x_{u,b}$ for each pair of business and student user, which can be though of as the probability $b$ is recommended to $u$. The value of a user-business recommendation is the expected number of activities a given user \textit{could} perform for a business which we will denote by $\al(u,b)$. Formally,
\
\begin{align*}
    \al(u,b) := \sum_{\text{Activities}} \I\{\text{Activity}\} \times \I\{\text{Business Type}\}, 
\end{align*}
\
\noindent where the indicators encode feasibility, i.e. $\I\{\text{Activity}\}  = 1$ if the activity is in the customers desired set and 0 otherwise, and $\I\{\text{Business Type}\} = 1$ if the business type is in the customers desired set and 0 otherwise. Thus $\al(u,b) \in \{0, 1, 2\}$ depending on the number of activities a user can perform with a business.  Finally, to ensure approximately balanced representation for all participating businesses we will require the ratio of the minimum expected number of impressions for a business to the maximum expected number of impressions for a business to be greater than or equal to 4/5. Putting it together we obtain the following recommendation LP, 
\begin{align*}
    \max_{\bx} & \sum_{u \in [U]}\sum_{b \in [B]} \al(u,b) x_{u,b} \\
    \text{s.t.} \,\, 
    & \sum_{b \in [B]} x_{u,b} = 1, && u \in [U] \\
    & \frac{4}{5}\left(\sum_{u \in [U]} x_{u,b}\right) \le \sum_{u \in [U]} x_{u,b'}, && b, b' \in [U], b \ne b' \\
    & x_{u,b} \ge 0.
\end{align*}

The objective function captures the total expected number of activities that could be performed. The first set of constraints ensures that the probabilities sum to 1 for each user. The second set of constraints ensures that, for any two businesses, the ratio of expected interactions for the first business over the expected interactions for the second business is at least $4/5$. The third set of constraints ensures nonnegativity for all probabilities.

Finally, the site uses these user specific probability distributions to generate the order of the businesses in the carousel.

\section{Omitted Proofs and Computations}\label{sec:appendix_pfs}

\subsection{Omitted Proofs from Section \ref{sec:badge_design}}


\begin{proof}[Proof of Proposition \ref{prop:badge_work_inc}]
Let $V \sim F$ be a positive-valued log-concave distribution with density $f$. First we will generalize Equation \ref{eq:opt_one} to $K$ thresholds. Recall $G_K(t_1, \ldots, t_K)$ is the expected number of activities performed in a system with $K$ badges and thresholds $\{t_i\}_1^K$. As users either hit the first threshold and redraw or leave the system, $G_K$ can be written as
\begin{align}\label{eq:opt_k}
   G_K(t_1, \ldots, t_K) &= \left(t_1 + G_{K-1}(t_2, \ldots, t_K)\right) \cF(t_1) + \E[V \I\{V < t_1\}].
\end{align}
Its derivative in $t_1$ is,
\begin{align}
   \frac{d}{dt_1}G_K(t_1, \ldots, t_K) &= \cF(t_1) - (t_1 + G_{K-1}(t_2, \ldots, t_K))f(t_1) \nonumber\\
   &\quad + \frac{d}{dt_1}\int_0^t v f(v) dv \nonumber \\
    &= \cF(t_1) - G_{K-1}(t_2, \ldots, t_K)f(t_1) \label{eq:deriv_k}
\end{align}
\noindent where the derivative follows from applying the fundamental theorem of calculus and simplifying. Note, if $\{t_i\}_1^K$ are maximizing for $G_K$, then it must be the case that $\{t_i\}_2^K$ are maximizing for $G_{K-1}$, and $\{t_i\}_3^K$ are maximizing for $G_{K-2}$ and so on. Now we will prove the three parts of Proposition \ref{prop:badge_work_inc}.

\textbf{(a)}
Since $F$ is log-concave, $\frac{\cF(t)}{f(t)}$ is non-increasing which implies that $G_K(t_1, \ldots, t_K)$ has a unique maximizer in $t_1$ at its critical point, $\frac{\cF(t_1)}{f(t_1)} =  G_{K-1}(t_2, \ldots, t_K)$. Thus $T_1$ is unique, and applying this observation iteratively we see that the optimal thresholds $T_1, T_2, \ldots, T_K$ are all unique, and solutions to the equations
\begin{equation}\label{eq:funct}
    \frac{\cF(T_i)}{f(t_i)} =  G_{K-i}(T_{i+1}, \ldots, T_K), \quad i \in [K-1].    
\end{equation}
Note the sequence of expected activities $\{G_{K-i}(T_{i}, \ldots, T_K)\}_{i=1}^K$ is decreasing since the thresholds used are all optimally chosen, and a feasible solution is using an empty threshold i.e.
\begin{align*}
G_{K-i}(T_{i+1}, \ldots, T_K) &= \max_t G_{K-i}(t, \ldots, T_K) \\
 &\ge G_{K-i}(0, \ldots, T_K) \\
 &= G_{K-i-1}(T_{i+2}, \ldots, t_K).
\end{align*}
Finally, recalling that $\frac{\cF(t)}{f(t)}$ is non-increasing by log-concavity, and since $T_i$ are solutions to Equation \ref{eq:funct}, the sequence $\{T_i\}$ is increasing. This completes (a).

\textbf{(b)}
By (a), for any $K$ and $i \in [K]$, $T_i \le T_K$ so we can simply compute $T_K$. Thus $T_K$, as it is the last threshold, is exactly the maximizer of Equation \ref{eq:opt_one} i.e., $T_K$ solves $\frac{\cF(T_i)}{f(t_i)} =  \E[V]$. Since $F$ is log-concave, its hazard rate $h(z) =  \frac{f(z)}{\cF(z)}$ is monotone and thus has an inverse. Rearranging gives the result.

\textbf{(c)}
Fix some $K \ge 3$ and let $\{T_i\}_1^K$ be the unique optimal thresholds for $G_K$. We will show that the marginal value of an additional threshold is decreasing by proving the increase in expected number of activities from going from $K-1$ optimal thresholds to $K$ is decreasing as $K$ increases.

Let $v(t) := \E[V\I\{V < t\}] + t\cF(t)$. Now rearranging Equation~\ref{eq:opt_k} yields
\begin{align*}
    & G_{K}(T_1, \ldots, T_{K}) - G_{K-1}(T_2, \ldots, T_{K}) \\
    & \qquad \qquad =  \E[V \I\{V < T_{1}\}] + T_{1}\cF(T_{1}) - F(T_{1})G_{K-1}(T_2, \ldots, T_{K}) \\
    & \qquad \qquad =   \max_t v(t) - F(t)G_{K-1}(T_2, \ldots, T_K)\\
    & \qquad \qquad \le \max_t v(t) - F(t)G_{K-2}(T_3, \ldots, T_K)\\
    & \qquad \qquad =   G_{K-1}(T_2, \ldots, T_K) - G_{K-2}(T_3, \ldots, T_K),
\end{align*}
where the first equality follows from the definition of $T_1$ and the first inequality follows from $G_{K-1}(T_2, \ldots, T_K) \ge G_{K-2}(T_3, \ldots, T_K)$.
\end{proof}

\subsection{Omitted Computation from Section \ref{sec:badge_design}}\label{sec:badge_compute}
In this section we will compute the optimal thresholds as described in simulation section of Section \ref{sec:badge_design}. First, we will iterate the expression Equation \ref{eq:opt_k} to obtain a closed form,
\begin{displaymath}
   G_i(T_i) = \E[V \I\{V < T_{i}\}] + \sum_{j=1}^i \left(\left(T_j + \E[V \I\{V < T_{j-1}\}]\right)\prod_{k=j}^i \cF(T_k)\right),
\end{displaymath}
where the second line follows from expanding out and rearranging the expression, and where we let $T_{K+1} = \infty$ for convenience. This expression allows us to easily recursively compute the optimal thresholds.

\textbf{($F = \text{Uniform}[0,5]$)}
We will compute the optimal thresholds and values starting from $T_3$. First,
$G_1(T_3) = \max_t (t + 2.5)\frac{5-t}{5} + \frac{t}{2}\frac{t}{5} = \frac{25}{8}$ and is maximized by $T_3 = \frac{5}{2}$. Next, $G_2(T_2,\frac{5}{2}) = \max_t (t + \frac{25}{8})\frac{5-t}{5} + \frac{t}{2}\frac{t}{5} = \frac{445}{128} \approx 3.48$ and is maximized by $T_2 = \frac{15}{8}$. Finally,
$G_3(T_1,\frac{15}{8},\frac{5}{2}) = \max_t (t +  \frac{445}{128})\frac{5-t}{5} + \frac{t}{2}\frac{t}{5} = \frac{121525}{32786} \approx 3.71$ and is maximized by $T_1 = \frac{195}{128}$. Rounding these numbers the optimal thresholds are $T_1 = 1, T_2 = 2$, and $T_3 = 3$.

\textbf{($F = \text{Uniform}[1,10]$)}
We will compute the optimal thresholds and values starting from $T_3$. First,
$G_1(T_3) = \max_t (t + 5.5)\frac{11-t}{10} + \frac{t+1}{2}\frac{t}{10} = \frac{157}{20}$ and is maximized by $T_3 = 6$. Next, $G_2(T_2,6) =\max_t (t + \frac{157}{20})\frac{11-t}{10} + \frac{t+1}{2}\frac{t}{10} = \frac{74409}{8000} \approx 9.36$ and is maximized by $T_2 = 3.65$. Finally,
$G_3(T_1,3.65,6) =\max_t (t + \frac{74409}{8000})\frac{11-t}{10} + \frac{t+1}{2}\frac{t}{10} \approx 10.47$ and is maximized by $T_1 = \frac{195}{128}$. Rounding these numbers the optimal thresholds are $T_1 = 2, T_2 = 4$, and $T_3 = 6$.

\newpage

\section{Omitted Figures}

\subsection{Badges}\label{sec:badge_figs}

\pagebreak

\subsection{Platform Highlights}\label{sec:site_figures}

\begin{figure}[ht]
    \centering
    \includegraphics[width=0.63\linewidth]{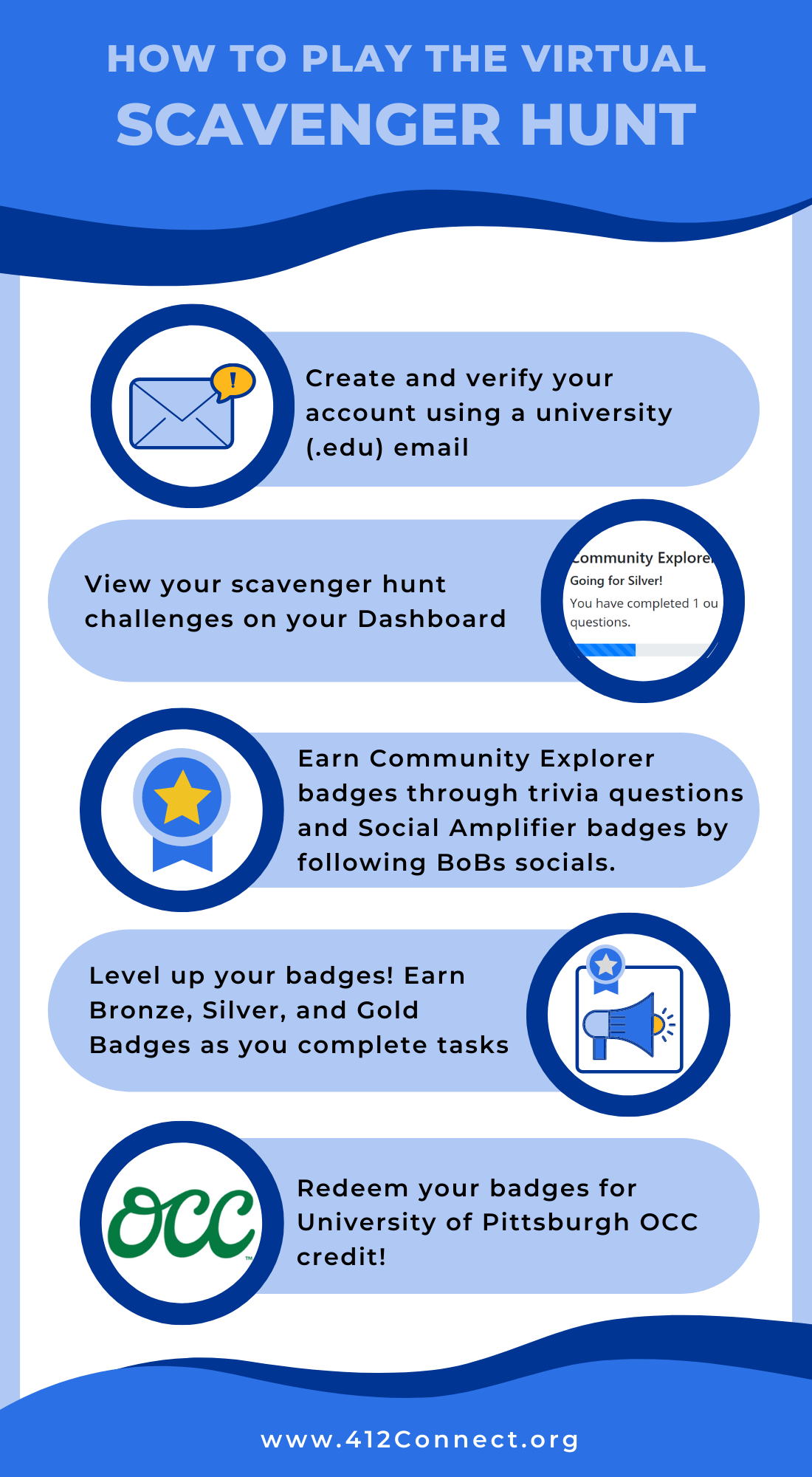}
    \caption{412Connect Instructions. {\rm A how-to guide for the 412Connect Platform starting from user sign-up and ending with the redemption of University of Pittsburgh OCC Credit}}
    \label{fig:how_to}
    \Description[How to use 412Connect.]{First, create and verify your account using a university (.edu) email. Second, view your scavenger hunt challenges on your Dashboard. Third, Earn Community Explorer badges through trivia questions and Social Amplifier badges by following Black-owned businesses' socials. Fourth, level up your badges! Earn Bronze, Silver, and Gold badges as you complete tasks. Finally, redeem your badges for University of Pittsburgh OCC credit!}
\end{figure}

\pagebreak

\begin{figure}[ht]
    \centering
    \includegraphics[width=0.36\linewidth]{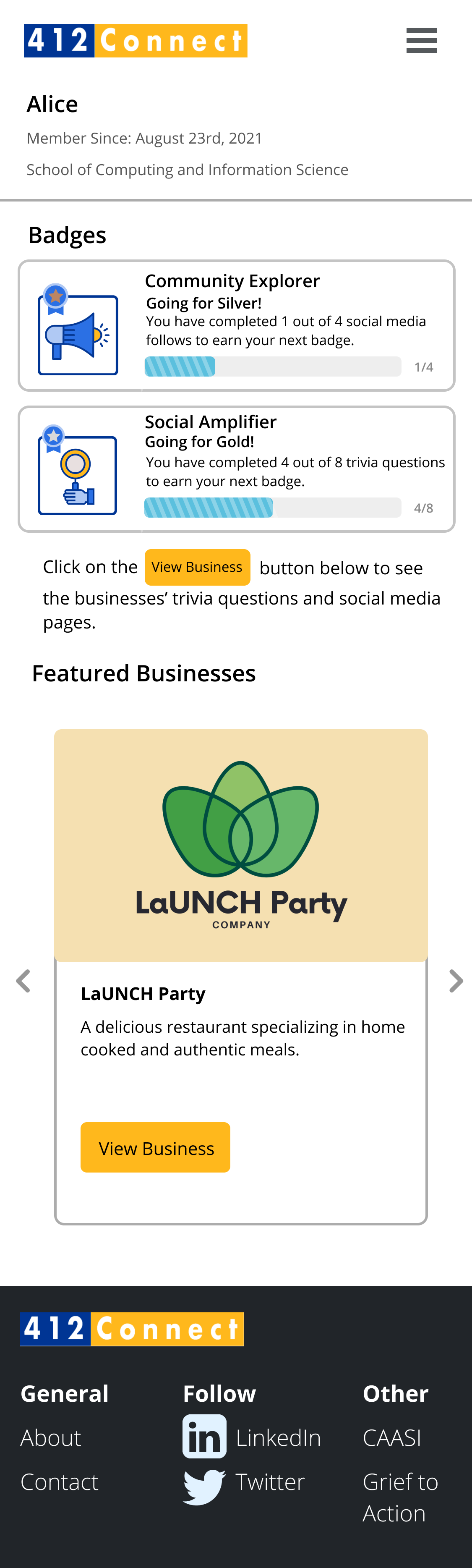}
    \caption{412Connect User Dashboard (Mobile Version). {\rm An example of the 412Connect User Dashboard. It contains the business carousel through which the user can peruse recommended businesses and navigate to a business page (Figure \ref{fig:wire_business}) by clicking ``View Business''. It also allows users to view their current progress in the area labelled ``Badges''.}}
    \label{fig:wire_dashboard}
    \Description[Example dashboard showing badges and business recommendations on mobile.]{At the top of the page the user name and affiliation are shown. Next, the currently earned badges are shown with progress bars to the next tier. At the bottom is the business carousel, showing a snapshot of one recommended business at a time.}
\end{figure}

\pagebreak

\begin{figure}[ht]
    \centering
    \includegraphics[width=0.42\linewidth]{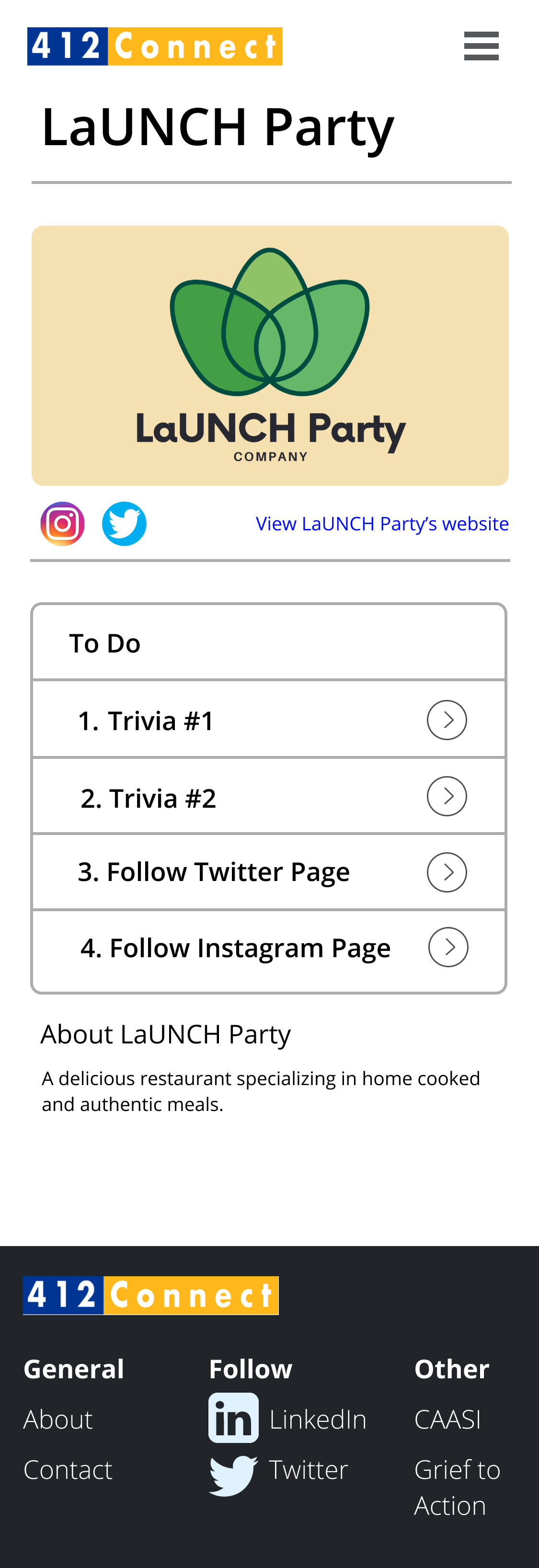}
    \caption{412Connect Business Page (Mobile Version). {\rm An example of a participating businesses page on the platform. In includes activities for students to complete as well as a business description and website and social media links for the business.}}
    \label{fig:wire_business}
    \Description[Example business page on mobile.]{At the top of the page the business's name and image are shown. Next are links to social media platforms and the business's website. Third is the to-do list of trivia and social media actions that can be taken. Last is the text description of the business.}
\end{figure}

\end{document}